\documentclass[a4paper,twocolumn,10.5pt,unpublished]{quantumarticle}
\pdfoutput=1
\usepackage[utf8]{inputenc}
\usepackage[english]{babel}
\usepackage[T1]{fontenc}
\usepackage{amsmath,amsthm,mathtools}
\usepackage{bm}
\usepackage{marvosym}
\usepackage{graphicx}
\usepackage{cite}
\usepackage{xcolor}
\usepackage{float}
\usepackage[title]{appendix}
\usepackage{hyperref}
\usepackage{physics} 
\usepackage{amsfonts}
\usepackage{tikz}
\usepackage{lipsum}
\usepackage{subcaption}

\begin{document}

\title{Deterministic Ans\"atze for the
 Measurement-based Variational Quantum Eigensolver}

\author{Anna Schroeder}
\orcid{0009-0006-4971-3265}
\affiliation{Merck KGaA, Darmstadt, Germany}
\affiliation{Department of Computer Science, Technical University of Darmstadt, Darmstadt, Germany}
\email{anna.schroeder@merckgroup.com}

\author{Matthias Heller}
\affiliation{Fraunhofer Institute for Computer Graphics Research IGD, Darmstadt, Germany}
\affiliation{Interactive Graphics Systems Group, Technical University of Darmstadt, Darmstadt, Germany}
\orcid{0000-0002-4774-5072}
\email{matthias.heller@igd.fraunhofer.de}
\author{Mariami Gachechiladze}
\affiliation{Department of Computer Science, Technical University of Darmstadt, Darmstadt,  Germany}
\email{mariami.gachechiladze@tu-darmstadt.de}
\maketitle

\begin{abstract}
 Measurement-based quantum computing (MBQC) is a promising approach to reducing circuit depth in noisy intermediate-scale quantum algorithms such as the Variational Quantum Eigensolver (VQE). Unlike gate-based computing, MBQC employs local measurements on a preprepared resource state, offering a trade-off between circuit depth and qubit count. Ensuring determinism is crucial to MBQC, particularly in the VQE context, as a lack of \textit{flow} in measurement patterns leads to evaluating the cost function at irrelevant locations. This study introduces MBVQE-ans\"atze that respect determinism and resemble the widely used problem-agnostic hardware-efficient VQE ansatz. We evaluate our approach using ideal simulations on the Schwinger Hamiltonian and $XY$-model and perform experiments on IBM hardware with an adaptive measurement capability. In our use case, we find that ensuring determinism works better via postselection than by adaptive measurements at the expense of increased sampling cost. Additionally, we propose an efficient MBQC-inspired method to prepare the resource state, specifically the cluster state, on hardware with heavy-hex connectivity, requiring a single measurement round, and implement this scheme on quantum computers with $27$ and $127$ qubits. We observe notable improvements for larger cluster states, although direct gate-based implementation achieves higher fidelity for smaller instances.
\end{abstract}

\section{Introduction}
The notion of quantum advantage is a highly intriguing application of quantum mechanics. Although it has been successfully showcased in computing problems tailored to a particular hardware~\cite{arute2019quantum,zhong2020quantum,bravyi2018quantum}, attaining computational speedup for problems with practical applications remains an important challenge. On the one hand, we already have Shor's algorithm~\cite{shor1994algorithms} and concepts from error-correction~\cite{campbell2017roads} to solve the factoring problem more efficiently than on any classical supercomputers, but such fault-tolerant schemes have stringent and potentially prohibitive requirements. On the other hand,  quantum devices available today are noisy, have a limited number of qubits, and can only operate short quantum circuits on which we can endeavor to solve practical problems. Therefore, the greatest challenge for the foreseeable future is figuring out how to utilize noisy and intermediate-scale quantum (NISQ) hardware.   

Hybrid quantum-classical algorithms and specifically the variational methods, which embed the
problems into parameterized short-depth quantum circuits and employ
the classical optimization routines to find the quantum
circuits that best solve the problem at hand, have attracted significant interest~\cite{moll2018quantum,mcclean2016theory,preskill2018quantum,peruzzo2014variational}. They address the problem of estimating the ground state energy of a quantum many-body Hamiltonian and have applications in quantum chemistry~\cite{tilly2022variational,peruzzo2014variational,amsler2023quantum}, high-energy physics~\cite{banuls2020simulating,kokail2019self,atas20212}, materials science~\cite{bauer2016hybrid}, and classical
optimization~\cite{bravo2019variational,borle2021quantum}. Alongside the success, this approach suffers from a few challenges: in general, 
training variational quantum algorithms is NP-hard~\cite{bittel2021training} and the error-mitigation might require a superpolynomial number of samples even for logarithmically shallow circuits, threatening any possible quantum advantage~\cite{quek2022exponentially}. Even though a few methods have been proposed to ease the former problem~\cite{bittel2022fast,zhou2020quantum,grimsley2019adaptive,grimsley2023adaptive,wierichs2020avoiding}, the latter looks more terminal. For significant classes of local Hamiltonians, any circuit aiming to prepare the ground state must have at least logarithmic depth~\cite{de2023limitations,bravyi2020obstacles,eldar2017local}, which poses the trade-off, one either resorts to inadequate approximation of the ground state, or the noise takes over the computation. In general, optimizing the depth of variational quantum algorithms is strongly QCMA-hard~\cite{bittel2022optimizing}. 

Here, we tackle the second problem. We propose a hardware-efficient protocol that borrows tools from measurement-based quantum computation~\cite{raussendorf2003measurement,briegel2009measurement}, enjoys extremely shallow circuits~\cite{broadbent2007parallelizing}, and is deterministic in measurement outcomes. It is well known that unlike dynamical languages, such as steps in the Turing machine or gates in the quantum circuit, in the measurement-based model, computation is performed just with local measurements on a many-body entangled state. While the circuit model of quantum computation defines its logical depth or computational time in terms of temporal gate sequences, MBQC allows a different temporal ordering and parallelization of universal logical gates. In this scheme, it is enough to achieve universality by analyzing the correlations that exist in many-body quantum states and, thus, is a very different, uniquely quantum way of thinking from all the dynamical computational models. Here, we study the possible advantages of this model of computation in variational algorithms. 

Several variants of measurement-based variational quantum eigensolvers (MBVQE) have been recently proposed~\cite{ferguson2021measurement, proietti2022native, qin2023applicability,chan2023hybrid, marqversen2023applications, majumder2023variational,huang2022near} (see also Appendix~\ref{app:overview} for an overview of previous works). However, they are either of probabilistic nature~\cite{ferguson2021measurement}, which causes exponential overhead or are not tailored to the current hardware capabilities requiring a high degree of qubit connectivity~\cite{chan2023hybrid, qin2023applicability}.  Here, we propose a general MBVQE scheme relying on the notion of quantum information flow defined for the one-way model to guarantee determinism independent of each measurement outcome~\cite{browne2007generalized,danos2006determinism}. Determinism is crucial to ensure that the variational steps in the classical training part converge even during the perfect noiseless simulations. It becomes even more pressing in the noisy regime when we have to incorporate current error-mitigation methods. From the deterministic models, we choose the specific ansatz architectures that are hardware efficient, and for two different examples, we study the trade-off between the size of the MB resource state translating into the number of parameters and the effectiveness of approximating the ground state. As a testbed, we run simulations for the Schwinger model and the $XY$-model and implement resource-conscious versions of our MBVQE ansatz of a four-qubit $XY$-Hamiltonian on an IBM quantum computer. We further study the efficacy of resource-state preparation on the heavy-hex-connected IBM quantum machines.

\section{Preliminaries}
\subsection{Variational Quantum Eigensolver}
The Variational Quantum Eigensolver (VQE) is a quantum-classical hybrid algorithm that approximates the lowest eigenvalue and its corresponding eigenvector of a given hermitian operator, typically the Hamiltonian of the investigated quantum system~\cite{peruzzo2014variational}. 
The basic idea of VQE is to generate quantum states using a parametrized quantum circuit, the so-called ansatz, and classically optimize the parameters~$\bm{\theta}$ such that the generated state~$\ket{\psi(\bm{\theta})}$ minimizes the cost function~$C(\bm{\theta})$, which is the expectation value of the Hamiltonian~$H$,
\begin{equation}
    C(\bm\theta) := \expval{H}{\psi{(\bm{\theta})}}\geq E_{gs} \;,
\end{equation} where $E_{gs}$ is the exact groundstate energy. 
Gradients of $\bm{\theta}$ can be efficiently measured directly from the quantum device using the parameter shift rule~\cite{mitarai2018quantum,schuld2018evaluating}, making them accessible for gradient-based optimizers. Various quantum ans\"atze, such as the problem-agnostic hardware-efficient ansatz~\cite{kandala2017hardware}, annealer-inspired QAOA~\cite{farhi2014a}, or the adaptively growing ADAPT-VQE~\cite{grimsley2019adaptive} have been proposed and extensively explored. The quantum chemistry community, in particular, has contributed numerous problem-inspired ans\"atze~\cite{anselmetti2021local}, many derived from the extensive literature on classical variational methods \cite{huggins2019a,romero2018strategies}.
Although there are known roadblocks to the productive use of VQE, such as the high counts of associated quantum measurements~\cite{gonthier2020measurements} or vanishing gradients (referred to as Barren plateaus)~\cite{mcclean2018barren}, VQEs remain the primary workhorse of quantum algorithms in the NISQ era.

\subsection{Measurement-based quantum computing}\label{sec:mbqc}

In MBQC, computation is carried out on a highly entangled resource state by local measurements only. 
The typical resource state is a graph state~\cite{hein2006entanglement}, however, other options have also been proposed~\cite{van2006universal,kissinger2019universal,gachechiladze2019changing}. A graph state $\ket{G}$ corresponds to a graph $G=(V,E)$, consisting of a set of vertices $V$ and edges $E$, in the following manner: Prepare $|V|$ qubits in the $\ket{+}$ state and associate them each to a vertex in $G$. Then apply the $CZ$~gate between two qubits if the corresponding graph vertices are connected:     
\begin{equation}
    \ket{G} = \prod_{\{i,j\} \in E} CZ_{i,j} \ket{+}^{\otimes|V|}.
    \label{eq:graphstate}
\end{equation}
Graph states can equivalently be defined by means of the stabilizer formalism. A graph state $\ket{G}$ corresponding to a graph $G=(V,E)$ is a unique eigenstate with the eigenvalue $+1$ to the following local Pauli stabilizers, 
\begin{equation}
    \ket{G} = X_i \bigotimes_{j \in \mathcal{N}(i)} Z_j \ket{G}, \quad \forall i\in V.
\end{equation}
Here each stabilizer operator is written for a vertex $i\in V$ and consists of the Pauli-$X$~gate on a vertex $i$ and Pauli-$Z$~gates acting on the graph-neighbors of $i$, denoted by the set $\mathcal{N}(i)$.
The most standard resource state for MBQC is a so-called cluster state; its associated graph is a regular two-dimensional grid. To harness universal quantum computation, the cluster resource state is paired with measurements in Pauli bases and particular measurement bases in $X-Y$ plane, parameterized by an angle $\theta$,  $\ket{\theta_{\pm}}:=\frac{\ket{0}\pm e^{i\theta} \ket{1}}{\sqrt{2}}$~\cite{raussendorf2003measurement}, which we denote sometimes as $R(\theta)$.
\begin{figure}
    \centering
    \includegraphics[width=0.9\linewidth]{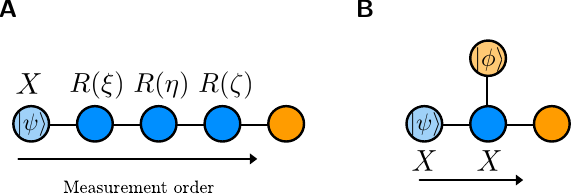}    
    \caption{\textbf{A}: A 5-qubit measurement pattern that implements an arbitrary single-qubit rotation $U_x(\xi) U_z (\eta) U_x (\zeta)$ from a linear graph state, and \textbf{B}: a 4-qubit pattern that performs a CNOT gate. Successive measurements of the blue qubits in the illustrated basis implement the desired quantum operation on the orange output qubit.}
    \label{fig:MBQCpatterns}
\end{figure}
Specific measurement patterns on snippets of the cluster state correspond to implementing quantum gates. See Fig.~\ref{fig:MBQCpatterns} for measurement patterns implementing an arbitrary single-qubit rotation and the CNOT gate. Typically, these patterns consist of input qubits, initialized to an arbitrary quantum state, output qubits, and qubits in between, depending on the pattern that is implemented. After measuring all qubits but the output qubits, the output qubits are modified to be equivalent to the input qubits with the intended gates applied. Due to the intrinsic randomness of quantum measurement outcomes, the output qubits generally carry additional random byproduct operators. The unique feature of MBQC is the ability to account for these byproducts and obtain a  deterministic output state on the fly.

To nurture an intuition of the MBQC framework, i.e., which operations are implemented with $R(\theta)$ basis measurements and how determinism is ensured, let us discuss a simple example that starts with a three-qubit, one-dimensional cluster state as a resource,
\begin{equation}
    \ket{G_3}: = CZ_{1,2}CZ_{2,3} \ket{\psi_\mathrm{in}}_1
    \otimes \ket{+}_2 \otimes \ket{+}_3\,,
\end{equation}
comprising qubits $q_1,q_2,q_3$, where the input qubit $q_1$ is initialized in an arbitrary state $\ket{\psi_\mathrm{in}} = \alpha \ket{0} + \beta \ket{1}$. 

Measuring $q_1$ in $R(\theta)$ basis  disentangles $q_1$ from the rest of the graph and yields one of the two possible subnormalized post-measurement states: 
\begin{align}
\begin{split}
   & \ketbra{\theta_\pm}_1 \ket{G_3}\\
   \propto & \ket{\theta_{\pm}}_1 CZ_{2,3} (\alpha \ket{+}_2 \pm e^{-i\theta}\beta \ket{-}_2) \otimes \ket{+}_3\\
    = &\ket{\theta_{\pm}}_1 \otimes 
    \ket{G_2^{\pm}},
\end{split}
\end{align}
where $ \ket{G_2^{+}}=CZ_{2,3} H_2 P_2(\theta) \ket{\psi_\mathrm{in}}_2 \otimes \ket{+}_3$  and $\ket{G_2^{-}}=CZ_{2,3} H_2 Z_2 P_2(\theta) \ket{\psi_\mathrm{in}}_2 \otimes \ket{+}_3$, where $H$ is the Hadamard gate and $P(\theta)$ is equivalent to rotation about the $Z$-axis up to a global phase $P(\theta) = e^{-i \theta/2}U_z(\theta)$. In other words, by projecting $q_1$ on $\ket{\theta_+}_1$, we effectively apply gates $H P(\theta)$ to the input state. Iterating this step and measuring $q_2$ in $R(\theta')$ basis, we  get the output qubit $q_3$ to be $HP(\theta')HP(\theta)\ket{\psi_\mathrm{in}}$ for two `$+$' outcomes on $q_1$ and $q_2$, but additional Pauli-$Z$ gates for other outcomes.

To obtain a deterministic post-measurement state, we need to account for the random byproducts. First, we take care of the ones appearing in $\ket{G^+_2}$ and $\ket{G^-_2}$. To this end, we use the relation between the measurement $\ket{\theta_{\pm}}$,
\begin{equation}
    \ketbra{\theta_-}_1 \ket{G_3} = \ket{\theta_-}\bra{\theta_+}_1 Z_1 \ket{G_3} \, .
\end{equation} 
Along with the graph stabilizer equation $Z_1\ket{G_3} = X_2 Z_3 \ket{G_3}$, we can shift that additional $Z_1$~gate onto the yet unmeasured qubits,
\begin{equation}
    \ketbra{\theta_-}_1 \ket{G_3} = \ket{\theta_-}\bra{\theta_+}_1 X_2 Z_3 \ket{G_3} \, .
\end{equation} 
This indicates that $\ket{G^-_2}$ is equivalent to $\ket{G^+_2}$ up to local Pauli byproducts. All Pauli-$Z$~byproducts that appear during measurements can be commuted similarly to the neighboring unmeasured qubits, introducing Pauli-$X$ byproducts. On the other hand, these new Pauli-$X$~byproducts can be accounted for through adaptive measurements as  $X \ketbra{\theta'_\pm}X=\ketbra{-\theta'_\pm} $, thus, the measurement basis has to be adapted from $R(\theta')$ to $R(-\theta')$.

\begin{figure*}[t]
    \centering \includegraphics[width=0.75\textwidth]{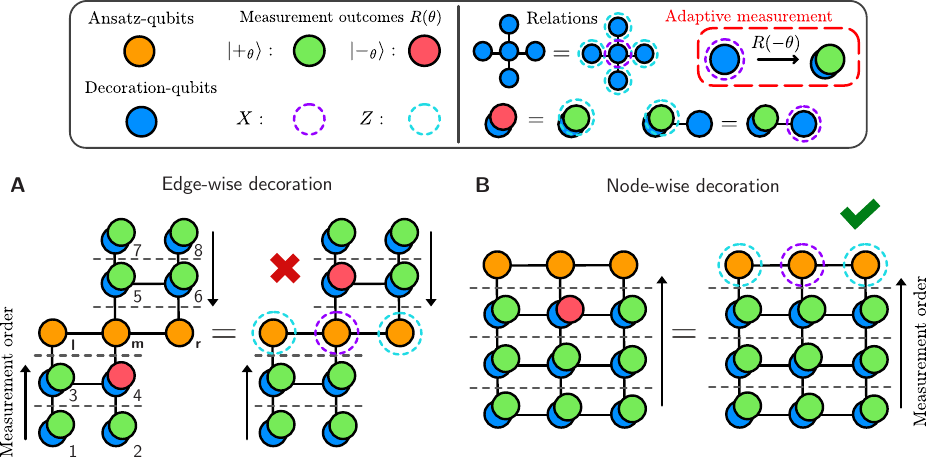}{\phantomsubcaption\label{fig:edgewise}
    \phantomsubcaption\label{fig:nodewise}}
    \caption{A graphical representation of MBQC and byproduct corrections. Orange qubits are called ansatz qubits in the MBVQE-context and are also the output qubits. The blue qubits are measured in varied bases $R(\theta_i)$. Measurements results are depicted with overlapping green and red circles, indicating a projection onto the positive or negative eigenstate, respectively. These eigenstates are related via Pauli-$Z$ operations which commute through the resource state via the graph state stabilizers. $X$ byproducts can be corrected via an adaptive measurement step defining a casual and temporal measurement order. Whether these byproducts can be corrected in MBVQE, depends on the selected decoration scheme. \textbf{A:} Ref.~\cite{ferguson2021measurement} introduced an edge-wise decoration scheme. Horizontally aligned qubits can be measured simultaneously. Here, two measurement orders coincide in the middle ansatz qubit and break the flow condition -- correcting for one byproduct introduces another on the other side. We cannot obtain deterministic outcomes. \textbf{B:} A deterministic decoration is obtained by layering the ansatz-graphs and connecting them vertically. Every $Z$ byproduct can be corrected with the stabilizer of the neighboring qubit in the subsequent measurement layer.  }
    \label{fig:detansatz}
\end{figure*}

To summarize, projections onto the negative eigenstates introduce $Z$~byproducts that can be accounted for by commuting them onto neighboring unmeasured qubits of the resource state. This introduces $Z$ and $X$~byproducts (specified by the stabilizers of the resource state) that can be, in the case of $Z$~byproducts, shifted further employing the stabilizer rules, and in the case of $X$~byproducts, compensated with adaptive measurements by swapping measurement bases $\theta'$ and $   -\theta'$ on the respective qubit. Byproducts acting on the output qubits can be corrected efficiently through classical post-processing. For a graphical illustration of these rules, see Fig.~\ref{fig:detansatz}. 
Note that adaptive measurements introduce a temporal and causal measurement order and a geometric condition for a ``correctable'' measurement pattern, known as the \textit{flow} in MBQC literature~\cite{danos2006determinism}. 
\subsection{The flow condition in MBQC}
The \textit{flow} condition was derived in Ref.~\cite{danos2006determinism} to characterize measurement patterns that can be executed deterministically by means of the byproduct corrections discussed in Section~\ref{sec:mbqc}. With the graph state as a resource and $R(\theta)$~basis measurements, determinism is ensured whenever the temporal measurement order, defined by the adaptive measurements, has a distinct flow without collisions. For intuitive understanding, let us study measurement patterns where the flow is missing (Fig.~\ref{fig:edgewise}) and present (Fig.~\ref{fig:nodewise}). In these examples, orange qubits are the output qubits, and blue qubits are measured layer by layer with the temporal order depicted by the arrow. Green and red overlapping circles indicate the measurement results of the $R(\theta)$~measurement. Thus, pictorially, the measurement pattern can be executed deterministically if it is possible to make all overlapping circles green. In Fig.~\ref{fig:edgewise}, two measurement orders coincide at the middle output qubit, which may cause a conflict in the byproduct correction. I.e., if only one of the qubits connected to the middle output qubit was projected onto the untargeted eigenstate $\ket{\theta_-}$, there is no way to account for this without causing an undesired phase kickback on the other side. For certain measurement outcomes in Fig.~\ref{fig:edgewise}, it is not possible to deterministically obtain a post-measurement state $\ket{\psi(\vec{\theta})}$ that does not depend on the signs of parameters $\pm \theta_i$. To be more precise, if the original resource state in Fig.~\ref{fig:edgewise} is denoted by $\ket{G_{11}}$, and after measuring qubits $4$ and $5$ we project on the outcomes $\ket{\theta_-}_4$ and $\ket{\theta_+}_5$, respectively, we have the equality with the case when we project on the outcomes $\ket{\theta_+}_4$ and $\ket{\theta_-}_5$,
\begin{align}\label{eq:pmoutcome}
\begin{split}
    \bra{\theta_-}_4 \bra{\theta_+}_5 \ket{G_{11}} =&\bra{\theta_+}_4 \bra{\theta_+}_5 Z_4\ket{G_{11}} \\
=&\bra{\theta_+}_4 \bra{\theta_+}_5  Z_l X_m Z_r Z_5 \ket{G_{11}} \\
=& \bra{\theta_+}_4 \bra{\theta_-}_5  Z_l X_m Z_r  \ket{G_{11}}.
\end{split}
\end{align}
On the other hand, if we project on the outcomes $\ket{\theta_+}_4$ and $\ket{\theta_+}_5$, we have the equality with the case when we project on the outcomes $\ket{\theta_-}_4$ and $\ket{\theta_-}_5$,

\begin{align}\label{eq:ppoutcome}
\begin{split}
    \bra{\theta_-}_4 \bra{\theta_-}_5 \ket{G_{11}} =&\bra{\theta_+}_4 \bra{\theta_+}_5 Z_4Z_5\ket{G_{11}} \\
=&\bra{\theta_+}_4 \bra{\theta_+}_5  Z_l X_m Z_r  \ket{G_{11}} \\
=& \bra{\theta_+}_4 \bra{\theta_+}_5  Z_l X_m Z_r  \ket{G_{11}}.
\end{split}
\end{align}
While the random byproduct on the left (l), middle (m), and right (r) qubits can be corrected in all the cases by classical post-processing, the post-measurement states are not necessarily equal in the two cases present in Eqs.~(\ref{eq:pmoutcome}, \ref{eq:ppoutcome}).
Consequently, any such protocol, including patterns without a flow, would require a postselection to obtain the ``all-green'' results, generally requiring an exponential sampling overhead. 

The pattern in Fig.~\ref{fig:nodewise} has a distinct measurement order and, thus, a flow. $\ket{\theta_-}_i$ outcomes on any blue qubits can always be compensated by the neighboring qubit in the upper measurement layer without introducing inequivalent possible post-measurement state parametrization.

\section{Measurement-based VQE}\label{sec:detmbvqe}

Ferguson et al.~\cite{ferguson2021measurement} presented two methods for transferring the gate-based VQE to MBQC. The first method involves translating the parametrized gate-based circuit into MBQC measurement patterns using the universal MBQC patterns introduced by Raussendorf et al.~\cite{raussendorf2003measurement}. 
While this approach may be suitable for platforms exclusively supporting MBQC, as anticipated for photonic quantum computers, it may be less appealing when evaluating the depth reduction performance of MBQC on gate-based machines. Certain ``cheap'' operations in the gate-based implementation, like single qubit rotations, require considerably many qubits and adaptive measurements to implement with MBQC.

The second MBVQE method introduced in Ref.~\cite{ferguson2021measurement} is a more ``MBQC-native'' approach, which forms the foundation of our work and inspired many other \cite{proietti2022native,qin2023applicability,chan2023hybrid,marqversen2023applications,majumder2023variational}. 
To illustrate, let us assume that we want to study an $n$-qubit Hamiltonian using MBVQE. In this method, we select $n$ output-qubits, a graph state, to act as the host for the parametrized $n$-qubit post-measurement state. These qubits are called \textit{ansatz qubits}. We have the flexibility to choose a specific graph structure of the ansatz qubits, aligning it with, for instance, known symmetries of the Hamiltonian or the qubit-connectivity of the hardware.
To introduce parametrization, we attach additional qubits called \textit{decoration qubits} to the ansatz qubits, all together forming a large graph state. These decoration qubits are measured in varied bases $R(\theta_i)$, introducing a variational parameter $\theta_i$ per decoration qubit.
Once all $k$ decoration qubits are measured, the state of the ansatz qubits depends on $k$ variational parameters. These parameters can be optimized classically, following the standard approach utilized in VQE.

A further freedom of design in this scheme is \textit{how} these decoration qubits are attached to the ansatz qubits and how many.

\subsection{MBVQE with edge-wise decoration}\label{subsec:edge-wise}

In Ref.~\cite{ferguson2021measurement}, the authors decided on an \textit{edge-wise} decoration strategy in the context of a problem-specific MBVQE ansatz for studying perturbed instances of the toric code. 
The ansatz qubits are chosen as the local-Clifford (LC) equivalent graph state of the noise-free toric code's ground state. Since the perturbative terms weaken the ground state's entanglement, decorations should allow tweaking the correlation strength of the ansatz qubits. Inspired by this idea, the edge-wise decoration suggests attaching four qubits to each edge in the ansatz-graph, with each of the two qubits from the added four, allocated to a vertex in an original graph (see Fig.~\ref{fig:edgewise}). 
While this MBVQE ansatz shows a favorable expressivity with respect to specific Hamiltonians \cite{ferguson2021measurement}, it is vital to acknowledge that this edge-wise decoration scheme is impractical whenever the ansatz graph state has a degree higher than one. Attaching multiple edge-wise decorations to a single ansatz vertex breaks the flow condition, prohibiting deterministic computation. 
\begin{figure}
    \centering
    \includegraphics[width=0.9\linewidth]{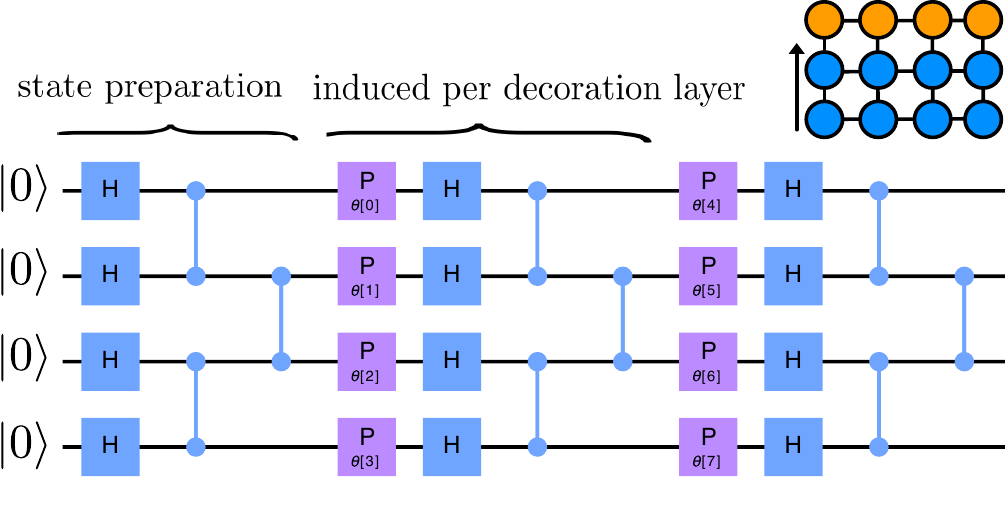}
    \caption{The effective quantum circuit that is deterministically implemented when all byproducts are corrected that may occur during measurements for the depicted MBVQE-ansatz. Increasing the number of decoration layers is equivalent to increasing the gate-depth.}
    \label{fig:nodewise_circuit}
\end{figure}
\subsection{Deterministic MBVQE with node-wise decoration}
In the context of MBVQE, a stringent treatment of random byproducts is essential when starting from a highly entangled graph state as a resource since the number of non-equalizable measurement outcomes grows exponentially with the number of flow-infringing qubits. This, naturally, makes postselection impractical with realistic problem sizes. To address this, we propose a decoration scheme that guarantees a flow in all cases.

To this end, we take a ``copy'' of the ansatz graph $G_0=(V_0,E_0)$, and call it the first decoration layer graph $G_1=(V_1, E_1)$. Every vertex $v^{(1)}_i\in G_1$ is then connected to a vertex $v^{(0)}_i\in G_0$. In an iterative way, we can then define the $l$-th decoration layer, where we add the $l$-th copy of $G_0$ and connect its vertices with the ones in the layer $(l-1)$. 
See Fig.~\ref{fig:nodewise} for a simple $3$-qubit example with $3$ decoration layers. Similar measurement patterns have been considered in Refs.~\cite{marqversen2023applications,majumder2023variational}. This approach allows the correction of any $Z$ byproducts by leveraging the stabilizer of the neighboring qubit in the above layer, starting the measurements from the bottom layer. We call such a decoration scheme \textit{node-wise decoration}. This decoration scheme differs from the edge-wise one in resource theoretic requirements: the edge-wise scheme maintains a planar graph, while the node-wise scheme requires an additional dimension to the ansatz graph. In Fig.~\ref{fig:nodewise}, our ansatz graph is linear, demanding the preparation of a regular 2D cluster state as the resource state. For a 2D ansatz graph, a three-dimensional cluster state becomes necessary, which can be used as a resource for fault-tolerant MBQC \cite{larsen2021a,raussendorf2005a,brown2018universal}. 

Since we aim to explore the depth-reduction capabilities of MBVQE compared to gate-based VQE, we choose resource states that can be efficiently prepared on current (superconducting) NISQ devices without the need for SWAP gates. Thus, we limit to MBVQE-patterns that can be generated from a two-dimensional cluster state. The gate-based equivalent circuit resulting from such patterns (see Fig.~\ref{fig:nodewise_circuit}) resembles a typical hardware-efficient entangler ansatz often used in NISQ applications~\cite{kandala2017hardware,moll2018quantum}. As entangler circuits typically contain nearest-neighbor interactions only, we demand that the resource-state preparation is efficient and preferably constant in gate-depth to justify an MB implementation on GB platforms for gate-depth reduction. 

While the cluster state topology is naturally given on some platforms like Google's Sycamore-chip ~\cite{arute2019quantum} with 2D-lattice qubit connectivity, other platforms may have sparser connectivity as IBM's quantum chips that exhibit a heavy-hex lattice \cite{hertzberg2020laser}. A naive implementation of the cluster state on such a device would require the usage of many SWAP gates. Fortunately, there are already known techniques to modify graph states with Pauli measurements and local unitaries ~\cite{anders2006fast,hein2006entanglement}. In Appendix~\ref{appendix:graph}, we demonstrate a measurement pattern that generates cluster states from heavy-hex graph states with Pauli measurements only. 

\section{Results and Experiments}
\subsection{Simulation of the $XY$- and Schwinger models}
\begin{figure*}[t]
 \centering
 \includegraphics[width = 0.8\textwidth]{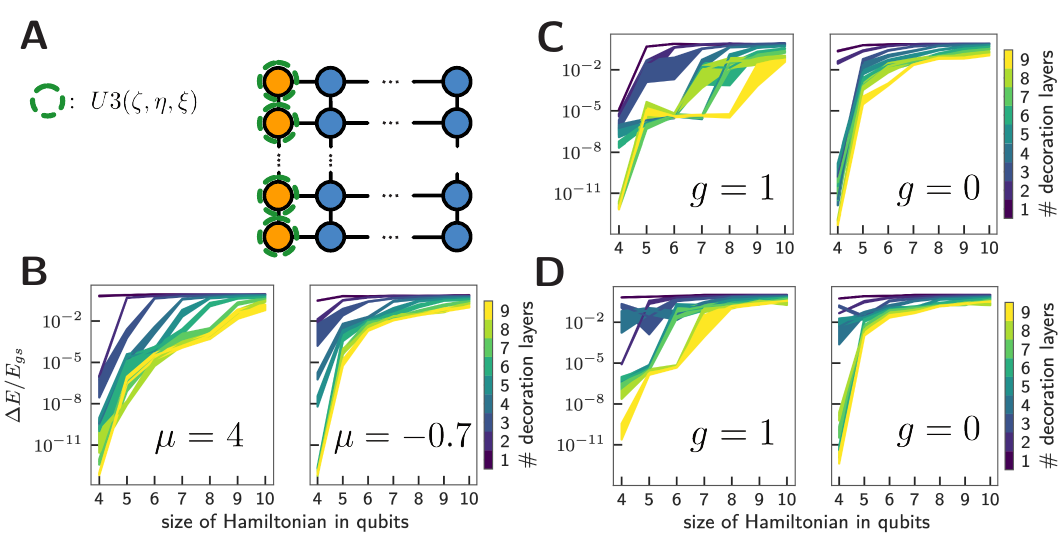}{\phantomsubcaption\label{fig:sim_ansatz}
 \phantomsubcaption\label{fig:schwinger_sim}
 \phantomsubcaption\label{fig:XY_scaling}
 \phantomsubcaption\label{fig:XY_scaling_wou}}
 \caption{\textbf{A:} In the scaling study, we vary the number of ansatz qubits corresponding to the problem dimension and the number of appended decoration layers, resulting in an $n\times l$ cluster state. Each decoration qubit introduces a variational parameter $\theta_i$. In addition to the variational decoration layer, we permit arbitrary local unitaries on the ansatz qubits following the measurement of the decoration qubits. \textbf{B:} The Jordan-Wigner transformed Schwinger model is examined for two to five fermions (corresponding to four to ten qubits) with $J = w = 1$. We investigate the Hamiltonian ground-state problem in two scenarios: an uncritical case ($\mu=4$, \textbf{left}) and a critical or close to critical setting  ($\mu=-0.7$, \textbf{right}). The relative errors of the approximate ground state energies compared to the exact solutions are plotted against the problem size in qubits. Line thickness indicates the variance from executing the same MBQVE circuit $10$ times, while the color represents the number of applied node-wise decoration layers. Increasing the layer count improves accuracy, but the scaling accessible in the current simulations is insufficient for the exact ground state recovery, in general. We could recover the exact ground state in the $4$-qubit scenario for both settings, but it becomes more difficult for higher qubit counts, although the estimate is more accurate in the uncritical case.
 \textbf{C:} This figure is analogous to Fig.~\ref{fig:schwinger_sim}, showing the results for the linear $XY$-model with an external Pauli-$Z$ perturbation of $d=0.01$ with problem sizes varied from four to ten qubits. We investigate two cases: the anisotropic case ($g=1$, \textbf{left}), equivalent to the Ising model, and the isotropic case ($g=0$, \textbf{right}). The observed trends are consistent with those in the previous figure, indicating that increased layering leads to improved approximations. Notably, the left case exposes a limitation of the ansatz, as disentangling a linear chain necessitates multiple layers of decorations. This can be avoided by modifying the ansatz qubits.
\textbf{D:} The same $XY$-model Hamiltonians are examined, but local $U3(\zeta,\eta,\xi)$ rotations on the ansatz qubits are omitted. This results in a slight quality decline, as the decoration layers are exhausted for implementing the local rotations. }
 \end{figure*}

We simulate the performance of the introduced MBVQE ansatz with two physical Hamiltonians, the Schwinger model~\cite{schwinger1951the,klco2018quantum},
\begin{align}
    H_S = & \frac{J}{2} \sum^{S-2}_{n=1} \sum^{S-1}_{k=n=1} (S-k) Z_n Z_k \nonumber\\
      +& w\sum^{S-1}_{n=1} (\sigma^{+}_{n}\sigma^{-}_{n+1} + \mathrm{H.C.})   + \frac{\mu}{2} \sum^{S}_{n=1} (-1)^{n} Z_n \nonumber \\ 
     -& \frac{J}{2} \sum^{S-1}_{n=1} n \ \mathrm{mod} \ 2 \sum^{n}_{k=1} Z_k 
     , 
\end{align}
where $\sigma^{\pm}_{n} = \frac{1}{2} (X_n \pm iY_n)$, and the perturbed $XY$-model,
 \begin{align}
     H_{XY}&=\sum_{i=1}^{n-1} \frac{1+g}{2} X_i X_{i+1}+\frac{1-g}{2}Y_i Y_{i+1}\nonumber\\
     &+ \sum_{i=1}^{2} d\; Z_i , 
 \end{align}
as test beds. 
An arbitrary Euler rotation
\begin{align}\begin{aligned}\\
\begin{split}U3(\zeta, \eta, \xi) = \begin{pmatrix} \cos\frac{\zeta}{2} & -e^{i\xi}\sin\frac{\zeta}{2} \\ e^{i\eta}\sin\frac{\zeta}{2} & e^{i(\eta+\xi)}\cos\frac{\zeta}{2}\end{pmatrix}\end{split}\end{aligned}\end{align}
is additionally applied to the output qubits, effectively implementing a basis change of both Hamiltonians. To study the efficacy of the final Euler rotations, we additionally test the MBVQE ansatz without them for $H_{XY}$.
We vary the size of both Hamiltonians from $4$ to $10$ qubits and employ $1$ to $9$ node-wise decoration layers. 
We execute the identical configuration ten times, generating the plots shown in Figs.~(\ref{fig:schwinger_sim}-\ref{fig:XY_scaling}), which display the average relative error between the approximate and the exact ground state, with the curve width representing the variance across the runs. The L-BFGS-B optimizer~\cite{liu1989on} paired with gradient estimates from finite differences are used to optimize the variational parameters.

Generally, the ground state is well approximated when dealing with small Hamiltonians. A linear increase in the number of layers improves the estimates but does not yield accurate solutions for larger Hamiltonians. This outcome was anticipated, as our ansatz emulates the hardware-efficient ansatz, exhibiting the same limitations.

The Schwinger model is examined at $\mu = -0.7$, critical in the second order\cite{kokail2019self,byrnes2002density}, and at $\mu = 4$, situated far from the critical point, in all cases with $J = w = 1$. The uncritical scenario is easier to approximate; the quality of the estimated ground state energy deteriorates around $9$ qubits. This deterioration is already evident for the $6$-qubit Hamiltonian in the critical case. See Fig.~\ref{fig:schwinger_sim}.

Our \textit{hardware-efficient} MBVQE ansatz demonstrates relatively better performance regarding the relative error than the entangler ansatz employed in Ref.~\cite{ferguson2021measurement}. In their approach, the relative error decreases below $0.2$ only using 8 decoration layers\cite{ferguson2021measurement}, whereas our ansatz achieves a significantly lower relative error of less than $10^{-5}$ already with two decoration layers.

Fig.~\ref{fig:XY_scaling} illustrates the results obtained from analyzing the $XY$-Hamiltonian. The Pauli-$Z$ perturbation term remains constant across all sites, with a fixed value $d=0.01$. Choosing $g=1$ corresponds to the transverse field Hamiltonian, while $g=0$ marks the critical point. In the case of the uncritical setting, where the ground state is a product state, we can achieve high levels of accuracy for larger Hamiltonians. However, as the system size increases, it becomes increasingly challenging to disentangle our initially highly entangled resource state with small numbers of decoration layers. For the critical $XY$-model, the results are similar to those of the Schwinger model. The accuracy of the estimates improves with an increasing number of decoration layers, but an additional number of layers accessible in the simulations is insufficient to reach the ground state accurately.

\subsection{$XY$-model on IBM hardware}
After discussing the optimal performance of the MBVQE entangler ansatz, our next objective is to gain a better understanding of its applicability on current gate-based quantum computers. 
Hardware experiments were carried out on the IBM platform, successfully executing quantum simulations with up to $127$ qubits, recently shown in Ref.~\cite{kim2023evidence}. While adaptive measurements, or \textit{dynamic circuits} as they are called within the \texttt{Qiskit} environment \cite{Qiskit}, are available, we found that in our use case they did not work as well as using a postselection strategy (of course, neglecting the sampling overhead) -- future improvements in the dynamical control of the hardware will change this.
Another challenge lies in the effective preparation of the resource state at the start. To address these concerns, we executed experiments using a measurement pattern that can be implemented on a heavy-hex lattice without requiring SWAP gates. Furthermore, this pattern is designed to be shallow regarding decoration layers and contains only up to one adaptive measurement.

\begin{figure*}[t]
    \centering
    \includegraphics[width = \textwidth]{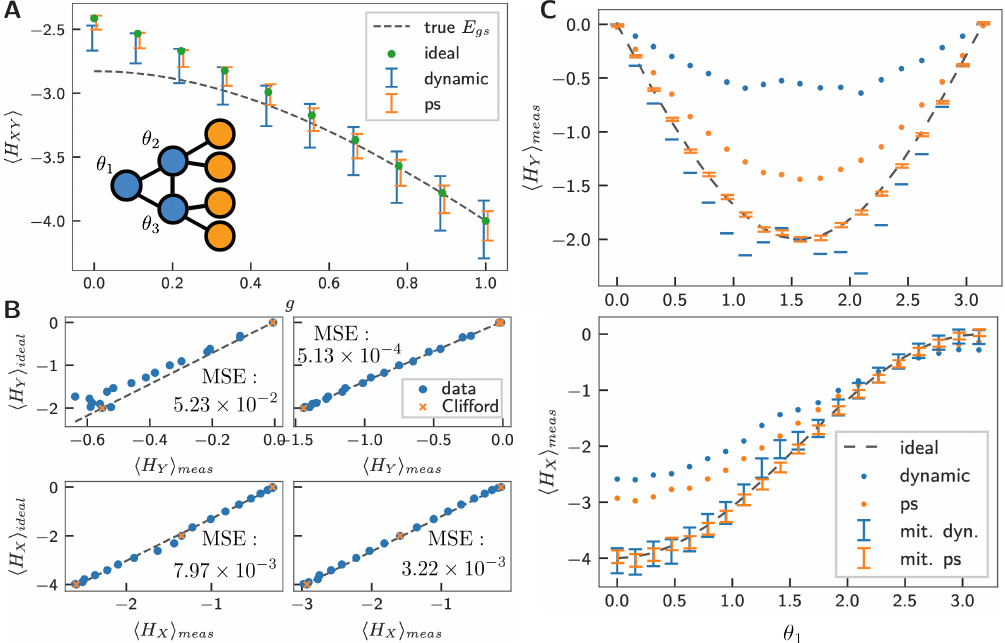}{
    \phantomsubcaption\label{fig:4_qubit_ansatz}
    \phantomsubcaption\label{fig:4_qubit_mitigation}
    \phantomsubcaption\label{fig:4_qubit_energy}
    }    \caption{\textbf{A:} We analyze the 4-qubit XY model using the depicted MBVQE-tree-ansatz on IBM hardware, employing a single variational parameter. This ansatz is suitable for larger values of $g$. The MBVQE procedure is performed twice, once using dynamic circuits (i.e., adaptive measurements) and once with postselection. Both error-mitigated estimates align closely with simulated data, falling within the expected error margin. \textbf{B:} Clifford data regression is conducted for dynamic circuits (left panels) and postselection (right panels) using $Y$ measurements (top) and $X$ measurements (bottom). We applied linear regression to the Clifford points and estimated the quality of the fits. Our analysis reveals that the mean-squared error (MSE) of the non-Clifford data points in relation to the fit is smaller for postselection. However, the fit fails to capture the error behavior of the $Y$ expectation value in the dynamic circuits. \textbf{C:} CDR successfully mitigates hardware noise, with the expected observation that errors in Y for the dynamic circuits are not captured. The error margins of the postselection method are more accurate. Combining data from the top and bottom, according to the Hamiltonian, and finding the minimizing parameters results in Plot A.}
    \label{fig:hwexp}
\end{figure*}

In order to accommodate the current hardware capacities, we propose a \textit{tree}-ansatz in Fig.~\ref{fig:4_qubit_ansatz}. Following the general correction scheme of the measurement byproducts in Section~\ref{sec:mbqc}, the only adaptive measurement we need to perform is that caused by a measurement outcome on the ``parent-qubit'' measured in $R(\theta_1)$ basis. We found that using more adaptive measurements leads to significantly higher degree of noise in all our use cases. Measurement outcomes of the remaining qubits can be treated by a classical post-processing of the measurement results on the output qubits.

We test the tree-ansatz for the periodic $4$-qubit $XY$-model, 
\begin{align}
    H_{XY} &\equiv-\frac{1+g}{2} H_X + \frac{1-g}{2} H_Y \nonumber\\
    &=-\frac{1+g}{2} (X_1X_2+X_2X_3+X_3X_4+X_4X_1)\nonumber\\
    &\quad+\frac{1-g}{2} (Y_1Y_2+Y_2Y_3+Y_3Y_4+Y_4Y_1).\label{eq:4_qubit_ham}
\end{align}
Note that, we introduce a minus sign in front of $H_X$ compared to the conventional $XY$-Hamiltonian.
However, our Hamiltonian is equivalent up to local unitaries to the conventional $XY$-Hamiltonian, since
$ Y_2 Y_4 H_X Y_2 Y_4 = -H_X$ and $Y_2 Y_4 H_Y Y_2 Y_4 =H_Y$.

We choose the $H_{XY}$ Hamiltonian, as for any $g$, it is straightforward to derive the expectation value for our ansatz,
\begin{align}\label{eq:variation_exp}
\begin{split}
    \expval{H_{XY}} =&-(1+g)(1+\cos \theta_1)\\
    -&\frac{1-g}{2}\left(\sin\theta_1(\sin\theta_2 +\sin \theta_3)\right),
    \end{split}
\end{align}
and minimize it for $\theta_2=\theta_3=\frac{\pi}{2}$. To say it otherwise, with our choice of the ansatz and the Hamiltonian, the measurements on the mid-layer qubits are always fixed in the Pauli-$Y$ basis. As a result, we are able to test dynamic circuits only varying one parameter, $\theta_1$ without demanding implementation of gradient optimization for other bases, which could possibly lead to additional error accumulation. In our hardware experiments, we chose to adapt the measurement basis of $\theta_2=\pm\frac{\pi}{2}$ depending on the outcome of the first measurement.

As an alternative approach to compare with, we use a postselection strategy, in which we only select those events for which the first qubit is projected on the $\ket{\theta_+}_1$. Then,
there is no need for any corrections in the second layer. Since for our choice of the ansatz and the Hamiltonian, we only need to implement Pauli-$Y$ measurements on the middle layer, we could effectively achieve the postselection with the classical post-processing, reducing the number of circuit runs. Furthermore, we use a combination of randomized compiling~\cite{wallman2015noise}, dynamical decoupling~\cite{PhysRevA.58.2733} and  Clifford data regression (CDR)~\cite{Czarnik2021errormitigation} (c.f.~Appendix~\ref{appendix:cdr} for more details) as error mitigation strategies.
We ran our experiments on \texttt{ibm\_algiers} ($27$ qubits) using a total budget of $100\; 000$ shots to estimate each expectation value.

We perform a scan over the parameter range $\theta_1 \in [0,\pi]$ using $20$ points in total, measuring the expectation values of $H_X$ and $H_Y$ (c.f.~Eq.~\eqref{eq:4_qubit_ham}) separately.
From these $20$ data points, we use the Clifford points $\theta_1=0,\,\pi/2\ {\rm and }\ \pi$ as mitigation points.
The measured values are different from the ideal classically simulated points due to noise in the circuit.
In Fig.~\ref{fig:4_qubit_mitigation}, we plot the measured values vs.~the ideal values, comparing the postselection strategy with the dynamic circuits for the $H_X$ and $H_Y$ expectation values.
Under the assumption of a global noise channel, there should be a linear relationship between the measured and ideal values~\cite{Czarnik2021errormitigation}.
As can be seen in Fig.~\ref{fig:4_qubit_mitigation}, this model works considerably better for the postselection strategy than for the dynamic circuits -- the mean-squared error of the recorded non-Clifford data points with respect to the CDR fit is larger for data points gathered from dynamic circuits.
We perform linear fits using the mitigation data and use them to mitigate the measured expectation values.

In Fig.~\ref{fig:4_qubit_energy}, we show the expectation values $H_X$ (upper panel) and $H_Y$ (lower panel). 
As can be seen, the mitigated values are in very good agreement with the ideal, simulated values. From the data shown in Fig.~\ref{fig:4_qubit_energy} we calculate the estimated groundstate energy of the Hamiltonian in Eq.~\eqref{eq:4_qubit_ham}

by adding the data of $H_X$ and $H_Y$ with the correct factors, see Fig.~\ref{fig:4_qubit_ansatz} for results.

\begin{figure}[h!]
    \centering
\includegraphics[width=0.9\linewidth]{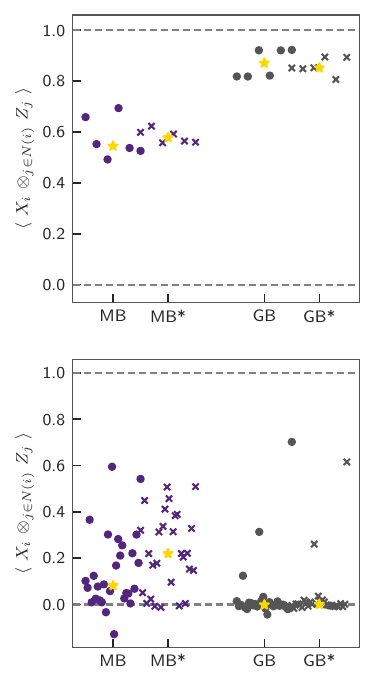}{
    \phantomsubcaption\label{fig:stateprep_small}
    \phantomsubcaption\label{fig:stateprep_big}
    }
    \caption{The performance of the measurement-based (MB) cluster state preparation is compared to the regular gate-based (GB) implementation, which requires high SWAP gate counts on the IBM hardware with heavy-hex connectivity. The plots display the expectation values of the cluster state stabilizer over 10,000 samples, with the median over all terms indicated by a star. Each method is evaluated once with read-out mitigation and once with additional randomization and dynamic decoupling, the latter marked by an asterisk.
    \textbf{Top:} In the case of the $2\times3$ cluster state, the GB method exhibits higher fidelity. This can be attributed to the absolute CNOT count, which is higher in the MB case, although the gate depth is smaller. Furthermore, the GB compiler can select the best-performing six qubits from the 27-qubit chip, while MB utilizes the entire chip. \textbf{Bottom:} The impact of depth reduction becomes evident for larger examples, such as the $4\times7$ cluster implemented on the 127-qubit machine. In GB, most expectation values collapse to 0, revealing total randomness. In contrast, MB shows significant signals, although still far away from the expected value of 1.}
    \label{fig:state_prep_2x3}
\end{figure}

\subsection{Cluster state preparation on IBM hardware}
In anticipation of advancements in adaptive measurements and decreasing readout errors, we develop a strategy to generate cluster states from heavy-hex graph states without the need for SWAP gates. To engineer this pattern, we leverage the graphical rules of Pauli measurements, which transform each ``honeycomb'' into a cluster square up to local byproduct gates, which can be corrected based on the measurement results on the auxiliary qubits. One drawback of this approach is the number of lost qubits. 

To assess the effectiveness of measurement-based resource generation, we perform experiments on a $27$- and a $127$-qubit quantum computer. Using the Pauli measurement-based preparation technique, we generate a $2\times3$ cluster state on the $27$-qubit machine and a $4\times7$ cluster state on the $127$-qubit machine.  As a point of comparison, we also generate these cluster states using a naive gate-based preparation method. 
Subsequently, we measure the stabilizers of the generated cluster states. The commuting stabilizer terms were grouped such that the observables to measure are reduced to two~\cite{toth2005entanglement}.

It should be noted that the measurement-based method induces local unitaries that modify the stabilizer of the state, which could be corrected using dynamic circuits. However, to avoid readout errors in such adaptive measurements, we refrained from correcting these unitaries to avoid additional sources of noise. Instead, we adjusted the stabilizers to be measured accordingly. For cluster qubits with an even number of neighbors, the additional unitaries are a combination from $\{I, Z\}$; both either commute with the graph stabilizer or introduce a minus sign that can be eliminated with classical post-processing. On the other hand, qubits with an odd number of neighbors could have additional Clifford gates from the set $\{I, Z, \sqrt{\pm iZ} \}$. These $\sqrt{\pm iZ}$ gates commute with $Z$ but transform $X$ to $\pm Y$. Consequently, we performed measurements in the $Y$ basis for such qubits whenever appropriate.

We used the \texttt{sampler} primitive provided in \texttt{ibm-runtime} to conduct our experiments.
The \texttt{sampler} has a built-in feature to correct read-out errors using matrix-free measurement mitigation as implemented in the \texttt{M3} package~\cite{PRXQuantum.2.040326}.
Furthermore, we implemented each experiment twice, once without additional modifications and another time using randomized compiling~\cite{wallman2015noise} and dynamic decoupling techniques~\cite{PhysRevA.58.2733}. The results show that these methods indeed reduce the scattering of data points, particularly in measurement-based implementations.
We ran the $27$-qubit experiment on \texttt{ibmq\_ehingen} and the $125$-qubit experiment on \texttt{ibm\_nazca}.
In all experiments, we used a total budget of $10\; 000$ shots to measure each of the two observables that had to be estimated to calculate the expectation values of all stabilizers.

We show the measured expectation value of each stabilizer state in Fig.~\ref{fig:state_prep_2x3}. 
In the case of the small six-qubit example, the naive gate-based method proved to be more effective. This can be attributed to the internal compiler selecting the best-performing $6$ qubits out of the available $27$ in the direct implementation. 
In contrast, the measurement-based method requires the use of the entire chip, even though some qubits may be unreliable due to bad calibration of two-qubit gates.
Additionally, although there is a clear advantage in terms of CNOT depth (defined as the CNOT count of the longest path in the compiled circuit as a directed acyclic graph) for the measurement-based method with a depth of $3$ compared to $11$ in the gate-based implementation, the total count of CNOT gates required in the circuit is $22$ for the measurement-based method and $16$ for the gate-based implementation.
However, the benefits of the measurement-based method become apparent in a larger example, such as creating a $4\times7$ cluster state using $125$ qubits. In this case, we are able to obtain the cluster state with a CNOT depth of $5$ and a total CNOT count of $142$. In comparison, the gate-based implementation results in a CNOT depth of $69$ and a total CNOT count of $186$. 
Although the stabilizer measurement in the measurement-based implementation still exhibits significant noise in most terms, we observe modest signals, especially with randomization and dynamic decoupling. In such cases, despite the overall resource state still being noisy and potentially impractical for MBQC, there is value in transitioning to the MBQC-based state preparation scheme.

\section{Conclusions and Outlook}

In this work, we focus on one of the most popular hybrid quantum-classical algorithms, the variational quantum eigensolver (VQE), in the measurement-based setting (MB). The MBVQE approach has recently gained a lot of attention. In our work, we put a particular focus on the importance of determinism when performing MBVQE, which we identify with the notion of flow in a resource state. For the cases where the flow is absent, we argue either for the inevitable exponential computational overhead caused by post-selection, or for the impossibility of convergence of the classical optimizer even in the perfect noiseless simulations. The issue becomes even more drastic if we need to use, for example, the read-out error mitigation.  

As a next step, we propose a new MBVQE protocol featuring a resource state that respects determinism and resembles the widely used problem-agnostic hardware-efficient VQE ansatz. Moreover, our ansatz can be prepared ``hardware efficiently'' on NISQ devices, such that it reduces the circuit-depth requirements by respecting the native qubit connectivity of hardware and avoids additional SWAP operations. We evaluate our approach using ideal simulations on the Schwinger Hamiltonian and $XY$-model of an increasing number of qubits. As a result, we obtain MBQC patterns that implement an instance of the entangler ansatz that gets more expressive by increasing the number of decoration layers in the proposed resource state. Additionally, we use IBM hardware, which allows for the mid-circuit measurements to test our results. 
Here, we propose a tree-shaped decoration scheme for the $4$-qubit $XY$-Model with a single adaptive measurement step and compare its performance in the case when determinism is ensured by post-selection (so no mid-circuit measurements). 
In this particular scenario  we find that the post-selection works significantly better, since the adaptive measurement steps induce additional hardware noise.

Finally, we propose an efficient MBQC-inspired method to prepare the resource state, specifically the cluster state, on hardware with heavy-hex connectivity, requiring just a single measurement round. We implemented this scheme on quantum computers with $27$ and $127$ qubits and observed notable improvements for larger cluster states, although direct gate-based implementation achieved higher fidelity for smaller instances.

From our analysis, we conclude that the MBVQE approach offers a fruitful test bed to characterize the expressivity of ansatz in VQE applications. As expected from the theory of MBQC, the MBVQE also offers a trade-off between different hardware requirements: e.g., a qubit count and gate-depth. Thus, depending on the architecture and bottlenecks of available near-term hardware, one can choose between the gate-based and MB-based approaches.

Recent announcements of next-generation quantum hardware, such as IBM Heron with native $CZ$ gates \cite{mckay2023benchmarking}, are promising for resource preparation in MBQC. It is also interesting to explore how such modular quantum hardware with classical interconnects behaves regarding resource state generation.
The effectiveness of MB approaches ultimately depends on the accuracy of adaptive measurements. Improvements in this area are anticipated on the path towards fault-tolerant quantum computing. Until these advancements materialize, it is worth exploring and adopting error-mitigation methods specific to MBQC \cite{gupta2023probabilistic}.

\begin{acknowledgments}
We thank Martin Kliesch, Wolfgang D\"ur, Luca Dellantonio, and Juani Bermejo-Vega for fruitful discussions. This work was in part supported by the research project \textit{Zentrum für Angewandtes Quantencomputing} (ZAQC), which is funded by the Hessian
Ministry for Digital Strategy and Innovation and the Hessian Ministry of Higher Education, Research and the Arts (MH). 
We acknowledge the use of IBM Quantum services for this work. The views expressed are those of the authors, and do not reflect the official policy or position of IBM or the IBM Quantum team.
\end{acknowledgments}

\onecolumn
\begin{appendices}
\section{Overview of other previous works}~\label{app:overview}

The work by Ferguson et al. \cite{ferguson2021measurement} inspired further investigations of the performance and resource requirements of other VQE instances in the MBQC framework. Proietti et al.~\cite{proietti2022native} studied an MBQC native way to implement the Quantum Approximate Optimization Algorithm (QAOA) \cite{farhi2014a}, which makes use of the fact that operations that are diagonal in the computational basis are exceptionally efficient in MBQC as ``input'' and  ``output'' qubits coincide \cite{browne2006one}. They tested the MBQC algorithm by simulating this method in the context of the k-max cut problem.
Qin et al. \cite{qin2023applicability} showed an MBQC version of the Hamiltonian variational ansatz \cite{wecker2015progress} that makes use of efficient 2-body operators implemented in the same sense as the QAOA operator. Combined with local rotations on the input/output qubits; it is possible to generate all Pauli-rotation.
Chan et al. \cite{chan2023hybrid} included these efficient MBQC-based multi-qubit Pauli-rotations, also referred to as Pauli-gadgets, within a gate-based circuit. This work included experiments on the IBM 7-qubit machine that indicate a possible advantage of operating gate-based hardware in MBQC manner.
Marqversen et al. \cite{marqversen2023applications} discuss and exhibit how MBQC can reduce hardware resources for applications such as VQE, demonstrate specific use cases, and introduce efficient ways to simulate MBQCs with tensor networks.
While Majumder et al. \cite{majumder2023variational} used almost the same MBQC pattern as in our work to generate their ansatz and acknowledge the importance of byproducts in MBVQE, but, instead of correcting them, they utilize this randomness to learn probability distributions better in the context of generative modeling. A part of these works used post-selection to guarantee the convergence of variational algorithm~\cite{ferguson2021measurement, proietti2022native}, while others used different measurement bases than $R(\theta)$ and achieved determinism~\cite{chan2023hybrid, huang2022near, marqversen2023applications}. While sufficient attention was not given to determinism in the initial works of MBVQE, recent papers (including ours) discuss the problem extensively for distinct settings and applications.

\section{Graph state modifications}\label{appendix:graph}
Here, we recall some valuable relations of graph states that are used in this work. Local  complementations (LC) implements \cite{adcock2020mapping,van2004graphical} the local unitary transformation of the following form:
\begin{equation}
 \ket{G_{LC}} = \sqrt{-iX_i} \bigotimes_{j \in \mathcal{N}(i)} \sqrt{iZ_j} \ket{G}.
\end{equation}

\begin{figure}
    \centering
    \includegraphics[width=0.8\linewidth]{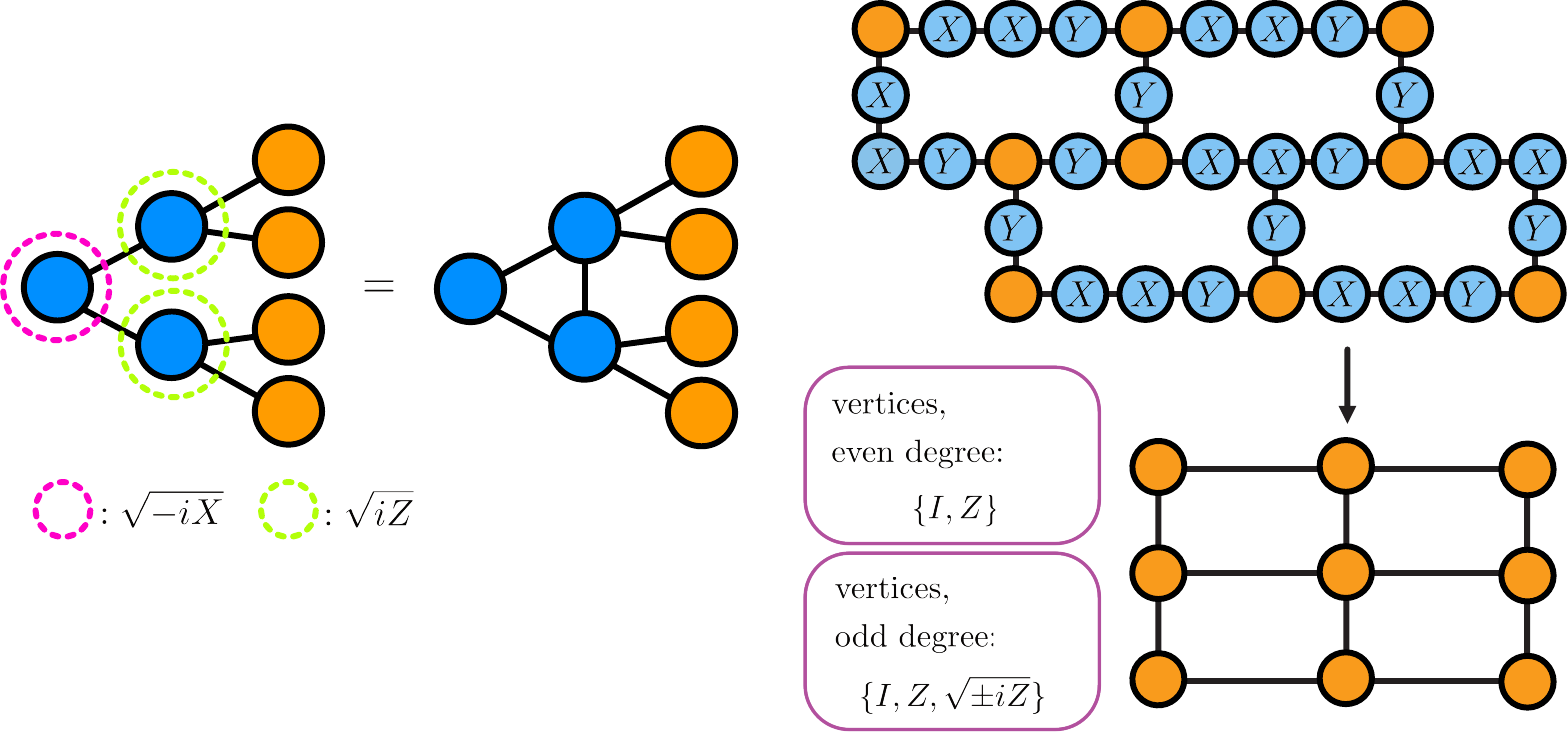}{
    \phantomsubcaption\label{fig:lclu}
    \phantomsubcaption\label{fig:measpattern}
    }
    \caption{\textbf{Left}: Graph states linked through local complementation are local Clifford equivalent. In our experiments, we choose to implement the left graph since it can be directly executed on the IBM device. \textbf{Right}: By preparing the heavy-hex graph state and measuring the blue qubits in the illustrated bases, it is possible to generate a cluster state without the need for SWAP gates. However, it's important to note that the measurement outcomes of Pauli measurements on graph states are not graph states but rather local unitary equivalents. Despite this, these ``byproduct'' operators can be efficiently computed from the measurement outcomes of the auxiliary qubits and accounted for in subsequent calculations due to Gottesman–Knill theorem.}
    \label{fig:lc}
\end{figure}

If we want to implement a specific graph state, we could choose to implement the representative from the LC equivalence class with the lowest edge count or the representative that best suits the hardware topology.
That is shown in Fig.~\ref{fig:lclu} -- the edge between the two qubits in the middle layer requires SWAP gates if implemented naively on quantum hardware with heavy-hex connectivity. However, it is easy to see that this edge can eliminated by local complementation on the parent-qubit. This graph can be directly mapped onto the respective hardware.

To generate a cluster state with constant gate depth from hardware featuring heavy-hex connectivity, we start by creating a heavy-hex graph state. 

It is known that graph states stay graph states up to specific local unitaries under Pauli measurements, and efficient graphical rules were developed to predict the post-measurement state \cite{hein2006entanglement, anders2006fast}. 
By using these rules and minding the commutation relation of measurement projectors and the emerging local unitaries (see Table \ref{tab:commutation_relations}), it is possible to design an effective measurement pattern illustrated in Fig.~\ref{fig:measpattern}. 

With this scheme, each \textit{honeycomb} becomes a square of the cluster state. Hence, on \texttt{ibmq\_ehingen}, we can generate a $2\times3$ cluster state, and on the 127-qubit machine \texttt{ibmq\_nazca} $4\times7$.

\begin{table}
    \centering
    \begin{tabular}{ccc}
         $P_{x,\pm} \sigma_z = \sigma_z P_{x,\mp}$, & \quad &  $P_{y,\pm} \sqrt{-i\sigma_z} = \sqrt{-i\sigma_z} P_{x,\pm}$, \\ 
         $P_{y,\pm} \sigma_z = \sigma_z P_{y,\mp}$, & \quad &  $P_{y,\pm} \sqrt{i\sigma_y} = \sqrt{i\sigma_y} P_{y,\pm}$, \\
         $P_{z,\pm} \sigma_z = \sigma_z P_{z,\pm}$, & \quad &  $P_{y,\pm} \sqrt{-i\sigma_y} = \sqrt{-i\sigma_y} P_{y,\pm}$, \\
         \quad & \quad &                                        $P_{y,\pm} \sqrt{i\sigma_z} = \sqrt{i\sigma_z}P_{x,\mp}$, \\
         
         \quad & \quad & \quad \\
         
         $P_{x,\pm} \sqrt{-i\sigma_z} = \sqrt{-i\sigma_z} P_{y,\mp}$,                & \quad &  $P_{z,\pm} \sqrt{-i\sigma_z} = \sqrt{-i\sigma_z} P_{z,\pm}$, \\
         $P_{x,\pm} \sqrt{i\sigma_y} = \sqrt{i\sigma_y} P_{z,\textcolor{red}{\mp}}$, & \quad &  $P_{z,\pm} \sqrt{i\sigma_y} = \sqrt{i\sigma_y} P_{x,\pm}$, \\
         $P_{x,\pm} \sqrt{-i\sigma_y} = \sqrt{-i\sigma_y} P_{z,\mp}$,                & \quad &  $P_{z,\pm} \sqrt{-i\sigma_y} = \sqrt{-i\sigma_y} P_{x,\textcolor{red}{\mp}}$, \\
         $P_{x,\pm} \sqrt{i\sigma_z} =\sqrt{i\sigma_z} P_{y,\mp}$,                   & \quad &  $P_{z,\pm} \sqrt{i\sigma_z} = \sqrt{i\sigma_z} P_{z,\pm}$. \\
    \end{tabular}
    \caption{The commutation relations play an essential role in simulating post-measurement states of multiple Pauli measurements on a graph state, as the resulting state is a graph state up to local Clifford operations. These additional operations alter successive measurements but in an efficiently simulable manner. The precise form of these operators and the table can be found in \cite{hein2006entanglement}. We opted to include the commutation relations in the appendix due to the presence of typos in the highlighted equations of the original source. }
    \label{tab:commutation_relations}
\end{table}

\section{Clifford-data regression}
\label{appendix:cdr}
Due to noise in the quantum hardware, expectation values extracted from circuit measurements have to be post-processed.
Several error-mitigation strategies exist, such as zero-noise-extrapolation~\cite{PhysRevLett.119.180509}, self-mitigation~\cite{ARahman:2022tkr}, or probabilistic error cancellation~\cite{van2023probabilistic,gupta2023probabilistic}.
For our experiments, we choose Clifford-data regression~\cite{Czarnik2021errormitigation} since dynamic circuits can be easily incorporated into this framework.

Given a parametrized circuit $\mathcal{C}(\theta_1,\dotsc \theta_n)$ and a Hamiltonian $H$, of which we try to estimate the expectation value $\bra{\psi}H\ket{\psi}$ at some given point $(\theta_1,\dotsc \theta_n)$, we first evaluate the circuit at near-Clifford points, i.e., we substitute some of the parameters by multiples of $\pi/2$, $\theta_i = n\cdot\pi/2,\ n\in \mathbb{N}$, for which the circuit can be simulated efficiently classically.
At these so-called regression points, we can find a relationship between the measured, noisy expectation values $\expval{H}{\psi}_{\rm noisy}$, and the exact, simulated ones,
\begin{equation}
    \expval{H}{\psi}_{\rm exact}=f(\expval{H}{\psi}_{\rm noisy},a_0,\dotsc a_n),
\end{equation}
where $a_i$ are model parameters.
In this work, we assume a linear relationship, i.e.,
\begin{equation}
    \expval{H}{\psi}_{\rm exact}=a_0 \expval{H}{\psi}_{\rm noisy}+a_1,
\end{equation}
which can be justified by the assumption of the existence of a global noise channel \cite{Czarnik2021errormitigation}.

\end{appendices}

\newpage
\twocolumngrid

\bibliographystyle{quantum}
\bibliography{sample.bib}

\end{document}